\documentclass[prl,aps,twocolumn,showpacs]{revtex4}
\usepackage{graphicx}
\begin{document}

\title{Inelastic light scattering from a Mott insulator}
\author{D. van Oosten,$^{1,2}$ D.~B.~M. Dickerscheid,$^{1,3}$ P. van der Straten,$^2$ and
        H.~T.~C. Stoof $^1$}
\affiliation{$^1$Institute for Theoretical Physics,
         University of Utrecht, Leuvenlaan 4, 
         3584 CE Utrecht,
         The Netherlands}
\affiliation{$^2$Debye Institute,
         University of Utrecht, Princetonplein 5, 
         3584 CC Utrecht,
         The Netherlands}
\affiliation{$^3$Lorentz Institute, Leiden University, P.O. Box 9506, 2300 RA Leiden, The Netherlands}

\date{\today}

\begin{abstract}
We propose to use Bragg spectroscopy to measure the excitation spectrum of 
the Mott insulator state of an atomic Bose gas in  an optical lattice.
We calculate the structure factor of the Mott insulator taking into account
both the selfenergy corrections of the atoms and the corresponding dressing of the 
atom-photon interaction.
We determine the scattering rate of photons in the stimulated Raman 
transition
and show that by 
measuring this scattering rate in an experiment, in particular the excitation gap of the 
Mott insulator
can be determined.
\end{abstract}
\maketitle

{\it Introduction.} --- A Bose-Einstein condensate in an optical lattice is a powerful new tool to
investigate strongly correlated Bose gases \cite{Duan2003a,Santos2004a}.
In particular, the experiment by
Greiner {\it et al.}~\cite{Greiner2002a} has shown that
 it is possible to achieve a quantum phase 
transition from a 
superfluid to a Mott-insulating phase by starting from an atomic Bose-Einstein condensate 
and loading this system into an optical lattice.
This phase transition was predicted to occur in the Bose-Hubbard model by 
Fisher {\it et al.}~\cite{Fisher1989a},
and Jaksch {\it et al.}~\cite{Jaksch1998a} were the first to make the crucial observation 
that the Bose-Hubbard model can be applied 
to bosonic atoms in an optical lattice. A mean-field theory that describes the two phases
of the Bose-Hubbard model was developed by van Oosten {\it et al.} \cite{vanOosten2001a}.

An important advantage of using atoms in an optical lattice to study the Bose-Hubbard
model, is that the system is free from disorder, which makes it possible to make very accurate 
predictions and measurements.
A good example of such a high-precision
measurement is Bragg spectroscopy. This technique has already been
used to coherently split a Bose-Einstein
condensate into two momentum components \cite{Kozuma1998a}, to 
measure the excitation spectrum of a trapped Bose-Einstein condensate 
\cite{Stamper-Kurn1999a},
and to measure the light-shifted energy levels of an atom in an optical lattice
\cite{Grynberg2001a}. 
Here we propose to use Bragg spectroscopy to measure the excitation spectrum of the
Mott-insulator state. In particular, one can in this way determine the value of the 
particle-hole gap in the 
excitation spectrum and study the behaviour of this gap as the system approaches the quantum 
critical point. Note that the excitation spectrum as obtained using Bragg spectroscopy will 
not yield what is generally referred to as the Mott gap, because this gap is associated with 
single-particle excitations. 
The value of the particle-hole gap is an interesting quantity in the study 
of quantum critical phenomena, but
it is also very important for the practical application of these systems in 
quantum information processing, since the gap determines the fidelity of the
Mott state.
It is important to realize that
the system of a Bose-Einstein condensate in an optical lattice is more complicated than
the above mentioned systems, because in this case many-body effects and strong correlations
have to be taken into account.
\begin{figure}[t]
\begin{tabular}{cc}
\includegraphics[width=4cm]{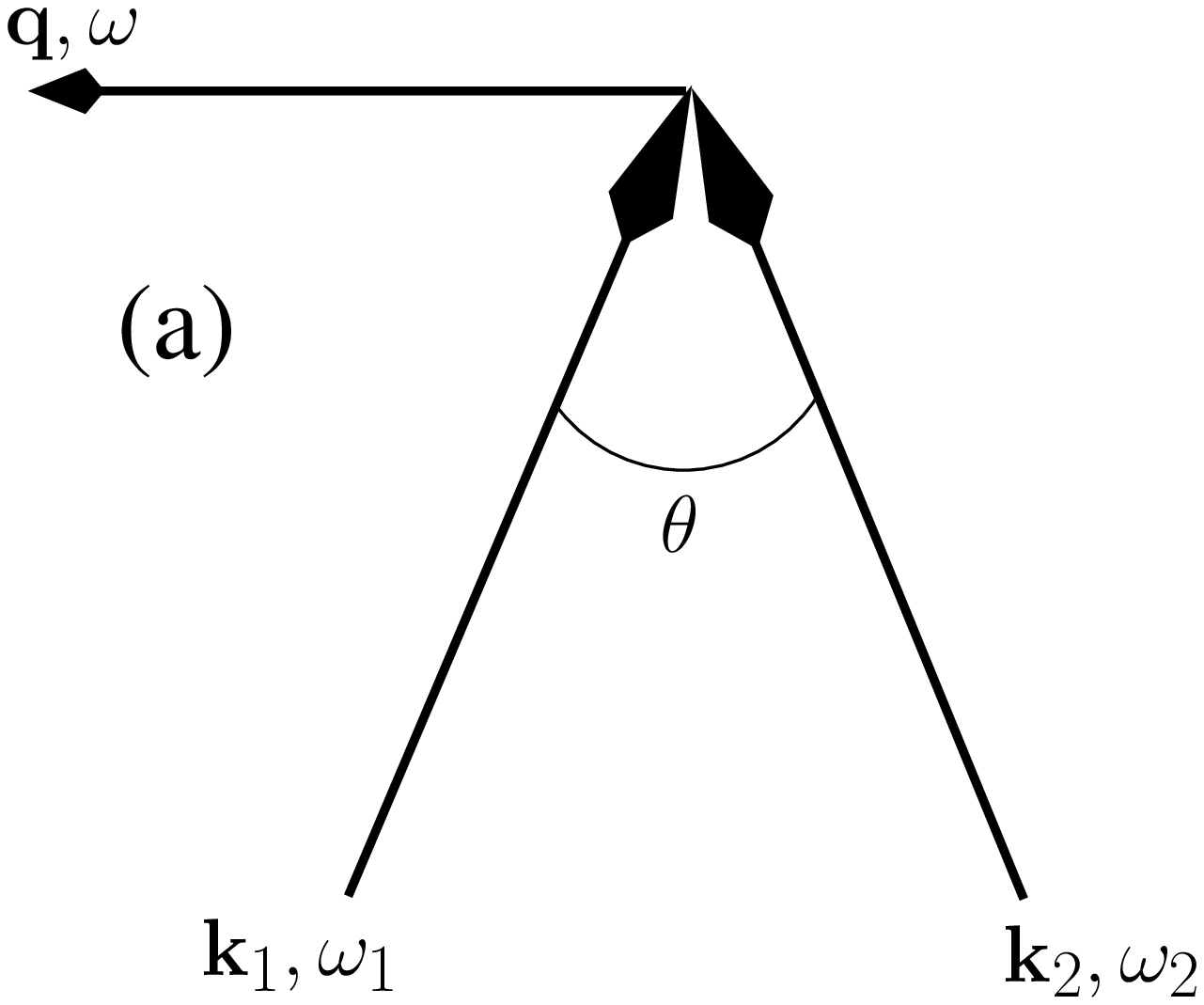}&
\includegraphics[width=5cm]{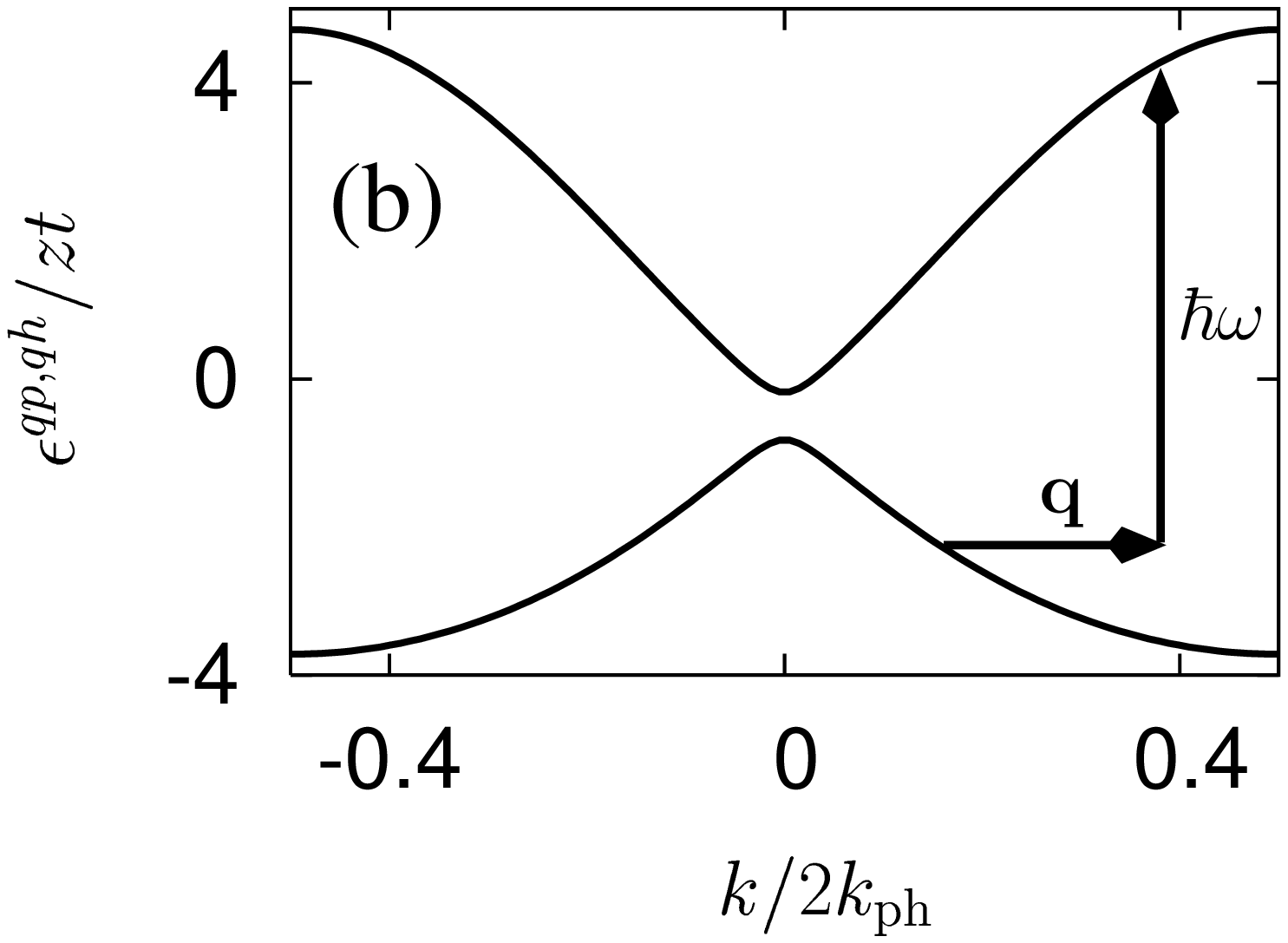}
\end{tabular}
\caption{(a) Setup for the proposed experiment.
(b) Particle and hole dispersions in units of the tunneling parameter
in a one dimensional lattice, for $U/zt=6$.
The horizontal arrow indicates absorption
of momentum, the vertical arrow absorption of energy.}
\label{fig:setup}
\end{figure}
In a Bragg spectroscopy experiment, two laser beams are used to make excitations in the 
system, as shown in 
Fig.~\ref{fig:setup}(a). 
The two lasers both have a large detuning
with respect to an optical transition in the atoms so that spontaneous emission is 
suppressed. However, the relative detuning can be very small.
When an atom absorbs a photon from beam two and is stimulated to emit a photon 
into beam one, the 
atom undergoes a change of momentum $\hbar{\bf q}=\hbar{\bf k}_2-\hbar{\bf k}_1$
and a change of energy $\hbar\omega=\hbar\omega_2-\hbar\omega_1$. 
In principle any optical transition could be used, but here we use the same
transition that is employed to create the lattice potential.
This means that the magnitude of the momentum is 
given by $\hbar q=2\hbar k_{\rm ph}\sin(\theta/2)$, where to a good approximation
$\hbar k_{\rm ph}=2\pi\hbar/\lambda$ 
is the photon momentum of both the lasers, $\lambda$ is equal to the wavelength  
of the lattice laser light
 and 
$\theta$ is the angle between the two laser beams.
By varying the angle between the two laser beams, any momentum 
between zero and $2\hbar k_{\rm ph}$ can be transferred and by varying the relative 
detuning between the beams, the amount of energy that is transferred to the system
can be controlled.

Calculating the scattering rate for a given momentum $\hbar{\bf q}$ and 
energy $\hbar\omega$ roughly speaking
involves counting the number of ways in which 
the requirements of momentum and energy conservation can be met.
To illustrate this process, we draw in Fig.~\ref{fig:setup}(b) 
the quasiparticle and quasihole dispersions
in the Mott insulator \cite{vanOosten2001a}, 
as is common in solid-state physics. The horizontal and vertical arrows in the figure indicate
the transfer of momentum and energy respectively.
Since energy is deposited in the system,
this scattering rate can be measured in a traploss experiment, or by determining
the increase in temperature of the atoms.

{\it The scattering rate.} --- To calculate the desired two-photon scattering rate 
we use Fermi's Golden Rule. In linear response, this can be expressed as
$
I({\bf q},\omega)=-2\mbox{Im}\left[ \Pi({\bf q},\omega) \right]/\hbar,
$
where $\Pi({\bf q},\omega)$ is the polarizibility of the medium.
The polarizibility can be written as $\Pi({\bf q},\omega) = 
\left( \hbar\Omega/2 \right)^2 \chi ({\bf q},\omega)$, with $\Omega$ the effective Rabi
frequency for the two-photon process and $\chi$ the susceptibility. The 
retarded susceptibility is given by
\begin{eqnarray}
\chi ({\bf q},\omega)&=&-\frac{V}{\hbar}\int d{\bf x} \int_0^\infty dt' 
e^{-i ({\bf q} \cdot {\bf x} -  \omega t')}\nonumber\\
&\times&\langle 
\left[ \hat{a}^\dagger({\bf x},t') \hat{a}({\bf x},t'),
\hat{a}^\dagger({\bf 0},0) \hat{a}({\bf 0},0)\right]\rangle,
\end{eqnarray}
with $V$ the volume and 
$\hat{a}^\dagger({\bf x},t')$ and $\hat{a}({\bf x},t')$ creation and annihilation
operators of the atoms.
Because the atoms are in an optical lattice, we can expand the field operators in terms of the 
Wannier
states of the lattice, which yields an expression in terms of creation and annihilation operators
for every lattice site. As mentioned previously, the Hamiltonian of the system 
then equals the 
Bose-Hubbard Hamiltonian with
a tunneling amplitude $t$, an on-site interaction energy $U$ and a chemical potential $\mu$.
Using the decoupling approach described in Ref.~\cite{vanOosten2001a}, 
we can write the atomic propagator in the Mott-insulator phase as
\begin{equation}
-\frac{1}{\hbar}G({\bf k},\omega)=\frac{Z({\bf k})}{-\hbar \omega^{+} 
+ \epsilon^{\rm qp}({\bf k})}
+\frac{1-Z({\bf k})}{-\hbar\omega^{+} + \epsilon^{\rm qh}({\bf k})},
\label{eqn:greens}
\end{equation} 
where the probabilities $Z({\bf k})$ and $1-Z({\bf k})$ account for the fact that an atomic
excitation contains both quasiparticle and quasihole contributions. The notation $\hbar \omega^{+}$
is shorthand for $\hbar\omega + i\xi$ with $\xi\downarrow0$.
The dispersions for the quasiparticle and quasihole excitations are given by:
\begin{eqnarray}
\epsilon^{\rm qp,qh}({\bf k})&=&-\mu + \frac{U}{2}(2N_0-1) + \frac{1}{2}\left(\epsilon({\bf k})
        \pm
\hbar\omega({\bf k})\right),
\end{eqnarray}
where $N_0$ is the filling fraction of the lattice and 
the function
$\epsilon({\bf k})=-t\sum_{j=1}^d \cos{2\pi k_j}$ corresponds to the lattice dispersion
in the experimentally relevant case of a regular square lattice.
The momentum $\hbar{\bf k}$ is here and from now on always written in units of $2\hbar k_{\rm ph}$, 
which means  that the first Brillioun zone runs from $k_j=-1/2$ to $1/2$.
The energy $\hbar\omega({\bf k})$ is given by 
$\hbar\omega({\bf k})=\sqrt{U^2+(4N_0+2)U\epsilon({\bf k})+\epsilon({\bf k})^2}$ 
and the probability $Z({\bf k})$ is given by $Z({\bf k})=(U(2N_0+1)+\epsilon({\bf k})+
			\hbar\omega({\bf k}))/2\hbar\omega({\bf k})$.
Using the Greens function in Eq.~(\ref{eqn:greens}), we find in first approximation
$\chi^0 ({\bf q},\omega)=t({\bf q})\left(\chi^0_+ ({\bf q},\omega) 
- \chi^0_+ (-{\bf q},-\omega)\right)$, 
where $t({\bf q})$ is a geometric factor that involves the appropriate overlap integral of 
the relevant Wannier functions (this will be discussed in Ref.~\cite{future}).  
Denoting integration over the first Brillouin zone as $\int_{1\rm BZ}$
,the contribution due to the creation of a particle-hole
pair is given by 
\begin{equation}
\chi^0_+ ({\bf q},\omega)=\frac{1}{2}\int_{1{\rm BZ}} d{\bf k} 
\frac{P({\bf k},{\bf k}+{\bf q},\omega)}
{-\hbar \omega^{+} + \epsilon^{\rm qp}({\bf k}+{\bf q})- \epsilon^{\rm qh}({\bf k}) },
\label{eq:chi}
\end{equation}
and the time-reverse process can be written as
$\chi^0_- ({\bf q},\omega)=\chi^0_+ (-{\bf q},-\omega)$.
This equation contains
the probability
$P({\bf k},{\bf k}+{\bf q},\omega)=\left( 1-Z({\bf k})\right)Z({\bf k}+{\bf q})$
for the creation of a hole with momentum ${\bf k}$ and 
a particle with momentum ${\bf k}+{\bf q}$, and an energy denominator that is associated with the 
energy cost $\epsilon^{\rm qp}({\bf k}+{\bf q})- \epsilon^{\rm qh}({\bf k})$ of that process.
This can readily be verified by taking the imaginary part of the susceptibility,
which is proportional to 
$\int_{1{\rm BZ}}d{\bf k}P({\bf k},{\bf k}+{\bf q},\omega)\delta(\hbar \omega-\epsilon^{\rm qp}({\bf k}+{\bf q})+\epsilon^{\rm qh}({\bf k}))$ and 
can be understood as Fermi's Golden Rule.
The actual computation of the above integral is too complicated to do analytically,
so that we have to resort to numerical methods. We achieve this by calculating the 
imaginary part of Eq.~(\ref{eq:chi}), which roughly corresponds to integrating over the surface 
in the Brillouin zone where the energy denominator vanishes. In practice, this amounts to numerically
finding the poles of the expression and determining their residue. 
The real part is calculated from the imaginary part using a Kramers-Kronig relation.

However, the results that one would obtain in this manner 
do not obey particle conservation. Physically, a Raman process with
momentum ${\bf q}$ couples to a density fluctuation $\rho({\bf q})$. For zero-momentum
transfer, $\rho({\bf 0})$ corresponds to the total number of particles and fluctuations are
impossible due to particle-number conservation. If we compute the imaginary part of 
Eq.~(\ref{eq:chi}) for ${\bf q}=0$ we find a spectrum which is nonzero, which means that
this approach is not sufficiently accurate.
\begin{figure}[ht]
\begin{tabular}{cc}
\includegraphics[width=4cm]{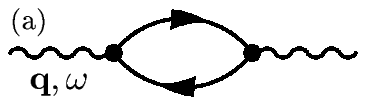}&
\includegraphics[width=4cm]{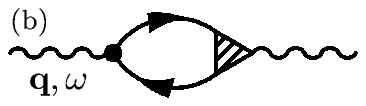}
\end{tabular}
\caption{Schematic representation of (a) Eq.~(\ref{eq:chi}) and
(b) Eq.~(\ref{eq:p_correct})}
\label{fig:feynman}
\end{figure}
The problem is due to the fact that 
in Eq.~(\ref{eqn:greens}) not the bare atomic propagator is used, but a dressed propagator
which contains a large self-energy correction given by
\begin{equation}
\hbar\Sigma({\bf k},\omega)=2 N_0 U +\frac{N_0(N_0+1)U^2}{\hbar\omega+U+\mu}.
\label{eqn:selfenergy}
\end{equation}
The first term on the right-hand side is the Hartree-Fock contribution, which is
also present in a Bose-Einstein condensate. The second contribution is due to
the correlations in the Mott insulator. 
Essentially this means that an atom moving through the Mott insulating background 
is dressed by all the other atoms.
As is known from quantum field theory \cite{Zinn-Justin}, one has to be careful when 
applying self-energy corrections to the calculation of the susceptibility,
because in general these corrections do not obey the required 
conservation laws (in this case particle-number conservation). Using field-theoretical 
methods,
we can derive so-called Ward identities, that show that every self-energy correction 
requires
a corresponding vertex correction in order to restore the conservation laws. 
Physically
this means, that if the atom is dressed, we also have to dress the atom-photon coupling.
Diagrammatically this is illustrated in Fig.~\ref{fig:feynman}.
This situation is very analogous to the situation in a superconductor, where the naive 
BCS calculation
of the electro-magnetic response is not gauge invariant and a more involved approach is 
needed \cite{schrieffer}. 
Using the relevant
Ward identity \cite{future} we can derive that the intuitive
probability function given above, has to be replaced by 
\begin{eqnarray}
P({\bf k},{\bf k}+{\bf q},\omega)&=&
\frac{2\hbar\omega - \epsilon^{\rm qp}({\bf k}+{\bf q}) + \epsilon^{\rm qh}({\bf k})}
{\hbar\omega({\bf k}+{\bf q})+\hbar\omega({\bf k})}\nonumber\\
&\times&
\left(Z({\bf k}+{\bf q})-Z({\bf k})\right).
\label{eq:p_correct}
\end{eqnarray}
Note that the probability now
vanishes when ${\bf q}\rightarrow 0$, so that particle conservation is indeed
no longer violated.  In fact, we can show that for small ${\bf q}$ and $\hbar\omega$ just above
threshold $P\propto q^2/\Delta_0^2$, where $\Delta_0$ is the gap for particle-hole
excitations.
\begin{figure}[hbt]
\includegraphics[width=8cm]{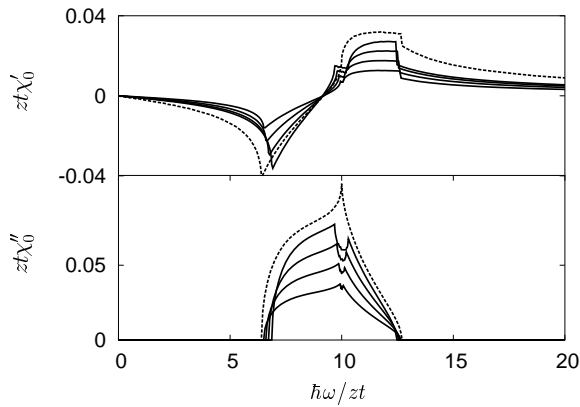}
\caption{Real and imaginary parts of the susceptibility for $U/zt=10$ and
${\bf q}=0.10,0.14,
0.18,0.20$
along a lattice direction, in two dimensions. The dotted line in the bottom figure
is the result
for ${\bf q}=0.001$ multiplied by 250 to show the behaviour for small $q$. }
\label{fig:spectrum2d}
\end{figure}

\begin{figure}[hb]
\includegraphics[width=8cm]{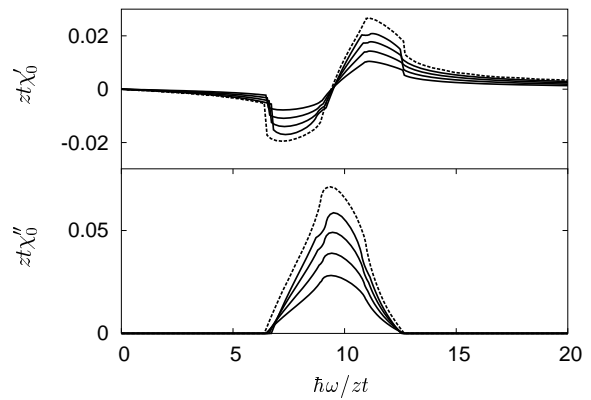}
\caption{Real and imaginary parts of the susceptibility for $U/zt=10$ and 
${\bf q}=0.10,0.14,
0.18,0.20$
along a lattice direction, in three dimensions. The dotted line in the bottom figure
is the result
for ${\bf q}=0.001$ multiplied by 250 to show the behaviour for small $q$. }
\label{fig:spectrum3d}
\end{figure}

{\it Results.} --- In Figs.~\ref{fig:spectrum2d} and \ref{fig:spectrum3d} the result
of a numerical integration is shown in two and three dimensions, respectively. 
Both calculations have been carried out for a regular square lattice and 
the momentum ${\bf q}$ is chosen a principal lattice direction. 
All energies in the following figures are given in units $zt$, where $z$ is the coordination
number of the lattice.

The imaginary part of Fig.~\ref{fig:spectrum2d} clearly shows singularities around 
$\hbar\omega=U$. These singularities are due to the fact that there are saddle points in the 
dispersion and that a saddle point in the dispersion causes an integrable singularity in the density
of states. These are so-called van Hove singularities \cite{vanHove1953a}.
It is
interesting to see, that the van Hove singularities split up as the momentum is increased, which is
caused by the fact that the saddle-point energy in the direction of ${\bf q}$ and the saddle-point energy
in the orthogonal direction(s) are shifted by different amounts. This is also visible in 
Fig.~\ref{fig:spectrum3d}. However, it is less clear in this case,
because the van Hove singularities  are more
smeared out in three dimensions. Also, the opening of the threshold for the two-photon
absorption in the three dimensional case is far less steep than in the
two-dimensional case.
To investigate possible collective modes in this system, we determined higher-order 
corrections
in the random-phase approximation (RPA). It can be shown that in RPA the susceptibility is given by 
$\chi ({\bf q},\omega)=\chi^0 ({\bf q},\omega)/(1-U \chi^0({\bf q},\omega)))$. 
This means that there is a resonance in the scattering rate
when the real part of $\chi^0({\bf q},\omega)$ is equal to $1/U$.
However, as can be seen from Figs.~\ref{fig:spectrum2d} and \ref{fig:spectrum3d}, 
the real parts in both cases are rather small compared to $1/U$
and in practice,  
including the RPA denominator does not qualitatively change our previous results.

\begin{figure}[hbtp]
\includegraphics[width=8cm]{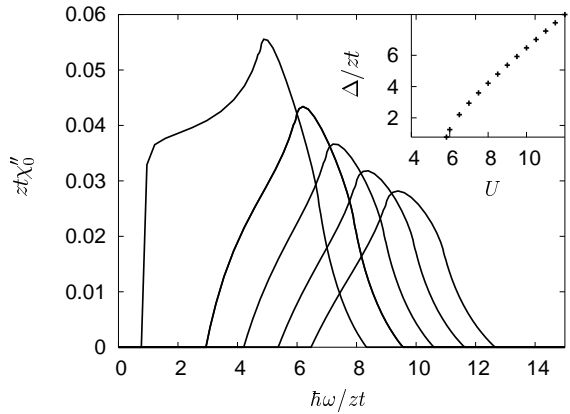}
\caption{Imaginary part of the susceptibility in a three dimensional lattice
for ${\bf q}=0.10$ along a lattice direction and $U/zt=5.83,7,8,9,10$
}
\label{fig:gap3d}
\end{figure}

In Fig.~\ref{fig:gap3d} we plot the imaginary part of $\chi_0$, 
for a range of values for the coupling constant $U/zt$,
and for a fixed momentum ${\bf q}=0.10$.
We see, that the threshold behaviour becomes steeper as we approach the critical
value of $U_c/zt\approx 5.83$. We also see that there
remains a nonzero gap when $U=U_c$. 
This is due to the fact that we are not considering a zero momentum 
excitation, due to the reasons given above. In the inset of Fig.~\ref{fig:gap3d}, we plot 
this gap $\Delta_q$ as a function of $U/zt$. For large $U$ the gap grows linearly with $U$
and for  $U$ close to $U_c$, the gap closes more rapidly. 
In the case of ${\bf q}={\bf 0}$ the gap would in our mean-field approximation close
as $\sqrt{U-U_c}$ when $U\downarrow U_c$, but for small nonzero ${\bf q}$ it closes as 
$\sqrt{U-U_c+\eta q^4}$, where the factor $\eta$ is a positive function of $U_c$ and $t$.

{\it Discussion.} --- In summary, we have proposed a means of studying Mott insulators in 
optical 
lattices, using the relatively well-known technique of Bragg spectroscopy. We have presented
spectra
that can be measured directly by trap loss or heating measurements. 
In a recent experiment by St\"{o}ferle {\it et al.} \cite{Stoferle2003a} the authors use a setup
where the laser beams are perfectly counterpropagating, which corresponds to a quasi-momentum
transfer of zero. As we have argued above, there should be no scattering in that case and
the signal can only be due to nonlinear response 
or to the fact that the system is inhomogeneous and of finite size.
We have found that by  measuring the threshold behaviour of the two-photon scattering rate 
at various
quasi-momenta, it is possible to determine the gap by extrapolation.
We have shown that for a theoretical description of Bragg spectroscopy on the Mott insulator
it is
absolutely essential to dress the photon-atom coupling, which is in a way unexpected, as the
corrections 
are zero in the case of an harmonically trapped gas. 
As a result it turns out that
although it is
common to use the language of solid-state physics to describe these systems, the physics is 
qualitatively very different due to the many-body effects.

\end{document}